\documentclass[preprint]{elsarticle}

\usepackage{amssymb}
\usepackage{amsfonts}
\usepackage{amsmath}
\usepackage{amsthm}
\usepackage{epsfig}
\usepackage{color} 
\usepackage{subfigure}
\usepackage{bbm}

\usepackage[english]{babel}

\begin{document}
\begin{frontmatter}
\title{Double-well chimeras in 2D lattice of chaotic bistable elements.}

\author{I.A. Shepelev}
\ead{igor\_sar@li.ru}
\address{Department of Physics, Saratov National Research State University,83 Astrakhanskaya Street, Saratov, 410012, Russia} 
\author{A.A. Bukh}
\ead{buh.andrey@yandex.ru}
\address{Department of Physics, Saratov National Research State University,83 Astrakhanskaya Street, Saratov, 410012, Russia} 
\author{T.E. Vadivasova}
\ead{vadivasovate@yandex.ru}
\address{Department of Physics, Saratov National Research State University,83 Astrakhanskaya Street, Saratov, 410012, Russia} 
\author{V.S. Anishchenko}
\ead{wadim@info.sgu.ru}
\address{Department of Physics, Saratov National Research State University,83 Astrakhanskaya Street, Saratov, 410012, Russia}
\author{A. Zakharova}
\ead{anna.zakharova@tu-berlin.de}
\address{Institut f\"{u}r Theoretische Physik, Technische Universit\"{a}t Berlin, Hardenbergstra\ss{}e 36, Berlin 10623, Germany}

\begin{abstract}

We investigate spatio-temporal dynamics of a 2D ensemble of nonlocally coupled chaotic cubic maps in a bistability regime. In particular, we perform a detailed study on the transition ``coherence -- incoherence'' for varying coupling strength for a fixed interaction radius. For the 2D ensemble we show the appearance of amplitude and phase chimera states previously reported for 1D ensembles of nonlocally coupled chaotic systems. Moreover, we uncover a novel type of chimera state, \textit{double-well chimera}, which occurs due to the interplay of the bistability of the local dynamics and the 2D ensemble structure. Additionally, we find double-well chimera behaviour for steady states which we call \textit{double-well chimera death}. A distinguishing feature of chimera patterns observed in the lattice is that they mainly combine clusters of different chimera types: phase, amplitude and double-well chimeras.

\end{abstract}

\end{frontmatter}

\maketitle

\section*{Introduction}
The study of the dynamics of multicomponent systems, such as nonlinear networks, ensembles of interacting nonlinear oscillators and distributed in space active systems is one of the most important directions in nonlinear dynamics. Interaction of nonlinear elements constituting complex systems results in a great variety of dynamical regimes and spatial structures. These questions are covered in monographs \cite{Kuramoto-1984, Afraimovich-1995, Mikhailov-1995, Landa-1996, Epstein-1998, Rosenblum-2001, Osipov-2007, Malchow-2007, Nekorkin-2012, SCH16b} and in many articles (for example, \cite{Kaneko-1989, Kaneko-1990, Nekorkin-1995, Nekorkin-1997, Nekorkin-1999, Belykh-2000, Belykh-2001, Pogromsky-2002, Belykh-2008, BIC11, Williams-2013, Pecora-2014}). In these and other works it is shown that one of the main features of nonlinear networks and spatially-organized active systems is the formation of patterns, such as synchronization clusters, spatial intermittency, steady state patterns, spatial chaos, various types of regular and chaotic wave processes, for example, spiral waves.

A new type of structures has been recently found: chimera states \cite{Kuramoto-2002, Abrams-2004, Abrams-2006, Panaggio-2015, SCH16b}. This structure is especially typical for ensembles of oscillators with nonlocal interactions. Chimera is a partial synchronization pattern consisting of coherent and incoherent clusters. Elements from the coherent cluster act in synchrony, while the oscillators from the incoherent cluster are not correlated and form a domain of spatial chaos with fixed borders. Chimeras were found in ensembles of phase oscillators \cite{Abrams-2008, Wolfrum-2011, Omel'chenko-2012, Xie-2014, Yeldesbay-2014, Maistrenko-2015}, periodic self-sustained oscillators \cite{Omelchenko-2013, Zakharova-2014, Sethia-2014, Omelchenko-2015a, Omelchenko-2015b, Vullings-2016, OME16}, chaotic oscillators and chaotic return maps \cite{Omelchenko-2011, Omelchenko-2012, Semenova-2015, Bogomolov-2016, Anishchenko-2016,Anishchenko-2016b, GHO16}, networks of oscillatory elements containing blocks of excitable elements \cite{ISE15b} or only excitable units \cite{SEM16}. Chimera structures were obtained not only in numerical simulations but also in experiments \cite{Tinsley-2012, Hagerstrom-2012, Smart-2012, Martens-2013, Kapitaniak-2014, Schmidt-2014}. Besides systems with nonlocal coupling chimera states were found for the case of global coupling \cite{Yeldesbay-2014, Sethia-2014, Schmidt-2014, Bohm-2014}, local coupling \cite{Laing-2015, Clerc-2016, Shepelev-2016} and also in single oscillators with delayed feedback where the delay time plays the role of virtual-space \cite{Large-2013, Large-2015, Semenov-2016}.

Chimeras in ensembles of chaotic oscillators significantly differ from chimera states in ensembles of phase oscillators and periodic self-sustained oscillators. Their emergence does not occur due to detuning of mean frequency of oscillators in incoherent clusters (which is typical for phase oscillators). Not any kind of chaotic behavior promotes chimera states. Apparently, hyperbolic chaos impedes the occurrence of chimeras \cite{Semenova-2015}. The models of chaotic systems (such as logistic map, R\"{o}ssler oscillator) demonstrating chimera states in networks of interacting units \cite{Omelchenko-2011, Omelchenko-2012, Semenova-2015, Bogomolov-2016, Anishchenko-2016, Anishchenko-2016b} are characterized by the regime of nonhyperbolic chaos, which occurs due to a cascade of period doubling bifurcations (Feigenbaum scenario \cite{Feigenbaum-1978}). The dynamics of ensembles of such elements with local interactions (for example, \cite{Kaneko-1990, Belykh-2000, Belykh-2001, Astakhov-1991, Anishchenko-boock}) shows a high level of multistability with a variety of coexisting states in the case of weak coupling. One can assume that nonlocal coupling makes chimera states more favorable. Therefore, in the present work we also focus on the nonlocal type of the interaction between the network elements.

Chaotic dynamics of a nonhyperbolic type is not restricted to the models of chaotic oscillators mentioned above. For instance, there are systems demonstrating bistability and the bifurcation of merging of chaotic attractors (such as Chua circuit \cite{Madan-1993} and other systems described in \cite{Sprott-2014}) for which chimera states have not been previously studied. The simplest model of a chaotic bistable system is a one-dimensional cubic map \cite{Anishchenko-boock, Scjolding-1983}. On the one hand, chaos in the cubic map is born through the Feigenbaum scenario as it is also the case for a logistic map. Therefore, one can assume that chimera states found in the ensembles of logistic maps with nonlocal coupling, should also be observed for networks of cubic maps. On the other hand, chaotic bistability and bifurcation of merging of chaotic attractors may introduce new features and lead to a novel type of chimera behavior. In this work we aim to investigate the patterns, and in particular chimera structures, occurring in networks of cubic maps and uncover their characteristic features related to the bistability of the chaotic behavior. 

Another question we address here is the impact of network topology on the occurrence of chimera patterns. Chimeras have been found for one-dimensional spatially-organized oscillatory ensembles with periodic boundary conditions (a ring) with nonlocal interaction, i.e. each element is coupled to a certain number of its nearest neighbors. There are also works on 2D-lattice \cite{Omelchenko-2012, Hagerstrom-2012, Schmidt-2014} and 3D-lattice \cite{Maistrenko-2015}. The chimeras in the form of spiral waves \cite{Martens-2010} have been observed in the model of 2D medium with phase dynamics of elements. In the works on chimeras in multidimensional ensembles a phase oscillator or its analog with discrete time \cite{Hagerstrom-2012} is used for the local dynamics. The occurrence of chimera patterns in 2D ensembles of elements with chaotic behavior remains to be understood. 

In the present work we investigate two-dimensional lattice of cubic maps with periodic boundary conditions and nonlocal interaction between the elements. The main question we address here is how the interplay of the bistability of the local dynamics and the 2D network structure influences the transition from coherence to incoherence. Moreover, we analyze the impact of the bifurcation of merging of chaotic attractors, which occurs in a single element, on the behavior of the network. In particular, we study the appearance of chimera patterns and uncover a novel type of chimera state which we call \textit{double-well chimera}.

\section{Model}

\begin{figure}[htbp]
\centering
\includegraphics[width=0.5\linewidth]{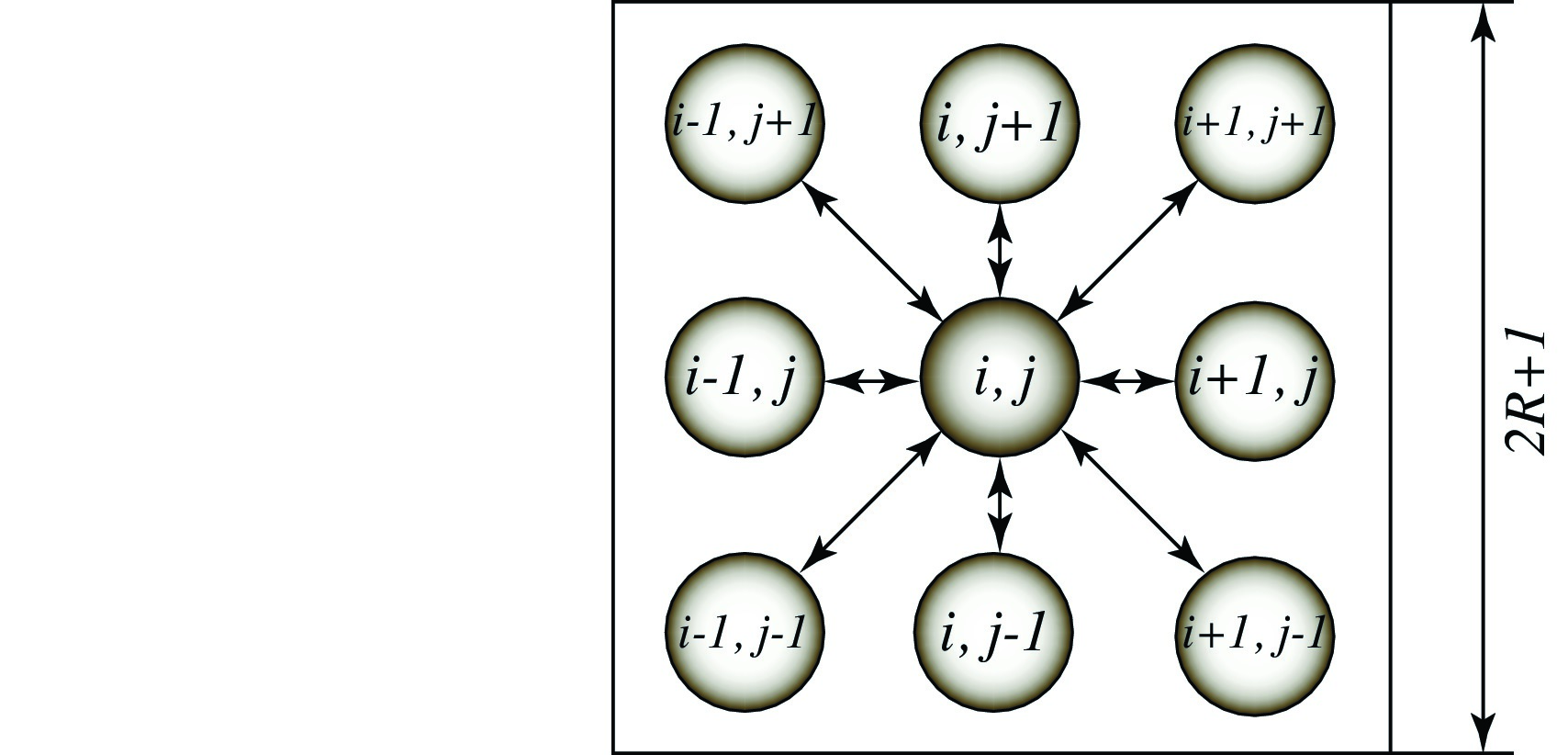}
\caption{Schematic representation of nearest-neighbor coupling in a 2D lattice.}
\label{pic:coupling_square}
\end{figure}

We study a 2D square lattice of nonlocally coupled cubic maps which is described by the following system of equations \eqref{eq:main}:
\begin{equation}
\begin{array}{l}
x_{i,j}(n+1) = f(x_{i,j}(n)) + \dfrac \sigma B
%Знак суммы
\sum\limits_{\tiny \begin{aligned}& k=i-R \\[-4pt] & p=j-R\end{aligned}}^{\tiny \begin{aligned}& i+R \\[-4pt] & j+R\end{aligned}}
%Выражение под знаком суммы
\left(f(x_{k,p}(n)) - f(x_{i,j}(n))\right),~~~i,j=1,...N,
\\
f(x) = \left(\alpha x - x^3\right)
\exp{\left[-\dfrac{x^2}{\beta}\right]}, ~~~ B=(1+2R)^2-1,
\\ 
x_{i+N,j}(n) = x_{i,j}(n), \quad x_{i,j+N}(n) = x_{i,j}(n),
\end{array}
\label{eq:main}
\end{equation}
where $i$ and $j$ specify the position of a lattice element and can be considered as discrete spatial coordinates ($X$ and $Y$), $n=0,1,2,...$ determines time, $f(x)$ defines the local dynamics and depends on the parameters $\alpha$ and $\beta$. The size of system under study is $N^2$, where $N$ is the number of elements in a one direction (along the $X$ or $Y$ axis). Boundary conditions are periodic in both directions. The interaction between the nodes of the network is nonlocal: each element of the lattice is coupled to its nearest neighbors located inside the square with the edge $2R+1$ and the element being located in the center of this square (see fig~.\ref{pic:coupling_square}). 
Thus, $B=(2R+1)^2-1$ is the total number of links for each element. Coupling strength is characterized by the parameter $\sigma$. For multidimensional ensembles a sphere with a specified radius $R$ is often considered instead of the square \cite{Omelchenko-2012, Hagerstrom-2012, Maistrenko-2015}. In this case $R$ is normalized by the total number of elements and is called coupling radius. Our numerical simulations show that for our particular problem both approaches give the same results (results not shown). At the same time, numerical algorithm is significantly simplified for the case of the squire and, therefore, the simulation time reduces essentially. Here we also use the normalization by the total number of elements $r=R/N$, where $r$ is the coupling radius and $R$ is a half of the square edge. In the present study the total number of elements is fixed $N^2=100\times 100$.

%
%%%%%%%%%%%%%%%%%%%%%%%%%%%%%% Fig.1 %%%%%%%%%%%%%%%%%%%%%%%%%%%%%%%%%%%%
%
%
%%%%%%%%%%%%%%%%%%%%%%%%%%%%%%%%%%%%%%%%%%%%%%%%%%%%%%%%%%%%%%%%%%%%%%%%

\begin{figure}[htbp]
\centering
\includegraphics[width=0.5\linewidth]{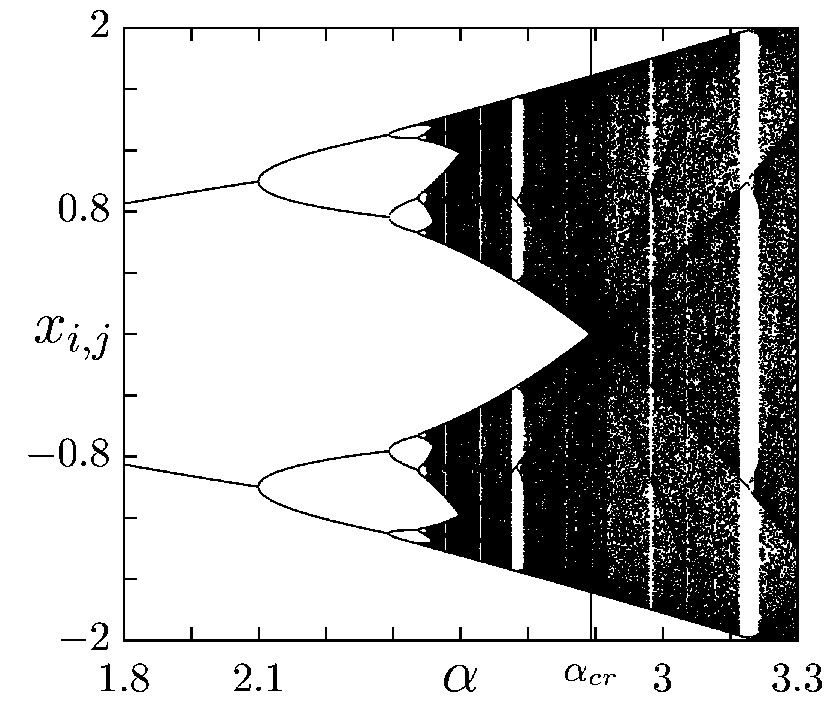}
\caption{Phase-parametric diagram of a single cubic map: variation of the parameter $\alpha$ for fixed $\beta=10$.}
\label{pic:diagram-alpha}
\end{figure}

The values of parameters $\alpha$ and $\beta$ of the cubic map are chosen within the regime of developed chaos. The cubic map demonstrates various dynamical regimes while varying the parameter $\alpha$ for the fixed value of $\beta=10$. For $\alpha<2.1$ the map is characterized by two fixed points. Each of them demonstrates the cascade of period doubling bifurcations with increasing $\alpha$ resulting in the appearance of two symmetric with respect to $x=0$ chaotic attractors as shown in fig.~\ref{pic:diagram-alpha}. The map becomes bistable for $\alpha<2.84$. Depending on initial conditions the values of $x_{n+1}$ are either positive or negative. Here we refer to these interval as positive and negative wells by analogy with an oscillator having double-well potential function. The single-well dynamics corresponds to the case when the oscillations occur in only one of the intervals of $x$: negative or positive. At $\alpha_{cr}=2.84$ the merging of chaotic attractors takes place. Therefore, for $\alpha>\alpha_{cr}$ all the phase points belong to the merged chaotic attractor and are distributed over both wells (a regime of monostability). This regime corresponds to double-well dynamics. The evolution of dynamical regimes for increasing parameter $\alpha$ is shown in a phase-parametric diagram (fig.~\ref{pic:diagram-alpha}).
%
%%%%%%%%%%%%%%%%%%%%%%%%%%%%%% Fig.2 %%%%%%%%%%%%%%%%%%%%%%%%%%%%%%%%%%%%
%
%
%%%%%%%%%%%%%%%%%%%%%%%%%%%%%%%%%%%%%%%%%%%%%%%%%%%%%%%%%%%%%%%%%%%%%%%%

In the case of ensemble of coupled maps the effective value of parameter $\alpha$ is changed. It explains the appearance of regular regimes and complex spatial structures during the transition from coherent chaos to the incoherent one for decreasing coupling strength \cite{Anishchenko-2016}. Indeed, the equation \eqref{eq:main} can be rewritten as follows:
\begin{equation}
\begin{array}{l}
x_{i,j}(n+1) = (1-\sigma) f(x_{i,j}(n)) + \dfrac \sigma B
\sum\limits_{\tiny \begin{aligned}& k=i-R \\[-4pt] & p=j-R\end{aligned}}^{\tiny \begin{aligned}& i+R \\[-4pt] & j+R\end{aligned}}
f(x_{k,p}(n)),
\\
f(x) = \left(\alpha x - x^3\right)
\exp{\left[-\dfrac{x^2}{\beta}\right]}, 
\end{array}
\label{eq:main2}
\end{equation}
In this case the multiplier ($1-\sigma$) occurring for the control parameter $\alpha$ decreases with increasing coupling strength $\sigma$. Thus, the lattice dynamics depends on the effective value of parameter $\alpha = \alpha_{eff} (\sigma)$. In the uncoupled case $\alpha_{eff}=\alpha$ and when coupling strength is strong $\sigma=1$ the system equations take the following form \eqref{eq:main3}: 
\begin{equation}
\begin{array}{l}
x_{i,j}(n+1) = \dfrac \sigma B
\sum\limits_{\tiny \begin{aligned}& k=i-R \\[-4pt] & p=j-R\end{aligned}}^{\tiny \begin{aligned}& i+R \\[-4pt] & j+R\end{aligned}}
f(x_{k,p}(n)).
\end{array}
\label{eq:main3}
\end{equation}
Here the term which is responsible for the local dynamics of the element disappears and all the elements perform forced oscillations being influenced by the neighboring nodes.
As a result, if we choose the value $\alpha$ corresponding to the merged chaotic attractor ($\alpha=3$) we observe the following situation. For the small coupling strength the system demonstrates the regime of merged chaotic oscillations in time. With increasing coupling strength the effective parameter $\alpha_{eff}$ decreases and the system switches to bistable regime. With further increasing coupling strength $\sigma$ the impact of neighbors is becoming predominant over the dynamics of an individual element. In this case we again observe the regime of merged chaos.

\section{Main dynamical regimes}

To get an overview over different patterns observed in system Eq. (\ref{eq:main}) we investigate the map of regimes in the plane of parameters $r$ and $\sigma$ for fixed $\alpha=3$ and $\beta=10$ (fig.~\ref{pic:diagram-r,sigma-chim}). The initial conditions used for the diagram are randomly distributed over the interval $0 \leq x_{i,j} \leq 1,~i,j =1,2, ... N$. Thus, all the elements of the lattice are located in the positive well for $t=0$. In the single-well regime these initial conditions lead to the oscillations only within the positive well. And in the regime of merged chaotic attractors the lattice elements are distributed between two wells even if initial conditions are chosen only in the positive well.

It is important to note that the system under study is characterized by high multistability and the borders between various dynamical regimes can be very narrow and can have a rather complex shape. The diagram shown in fig.~\ref{pic:diagram-r,sigma-chim} illustrates the main types of spatio-temporal dynamics of the system which occur by varying the coupling strength $\sigma$ and the coupling range $r$. Due to high multistability, different regimes in system Eq. (\ref{eq:main}) overlap. One can distinguish between double-well and single-well spatial patterns (in fig.~\ref{pic:diagram-r,sigma-chim} regions I and II, respectively). The main dynamical regimes observed for the lattice Eq. (\ref{eq:main}) are full chaotic synchronization, partial coherence (oscillating synchronization) according to the terminology used in \cite{Anishchenko-2016b}, chimera states and spatial incoherence (in fig.~\ref{pic:diagram-r,sigma-chim} A, B, C and D, respectively).
We find the regimes characterized by different distributions of the instantaneous states of lattice elements over the two wells which correspond to positive and negative value intervals (regions I, II and II in fig.~\ref{pic:diagram-r,sigma-chim}). For strong coupling strength the double-well spatial structure is dominating: instantaneous states of the elements are distributed between two wells and the instantaneous values of $x_{i,j}$ can be both positive and negative (region I in fig.~\ref{pic:diagram-r,sigma-chim}). In more detail, the dynamics in time in this case is as follows: the oscillations of an individual element can be either within one well or demonstrate switching between the wells. On the contrary, for the single-well structures all the lattice elements oscillate in one and the same well: the values $x_{i,j}$ are either positive or negative (region II in fig.~\ref{pic:diagram-r,sigma-chim}). 
For the large values of both coupling parameters $\sigma$ and $r$ we observe synchronization of chaotic switching (region III in fig.~\ref{pic:diagram-r,sigma-chim}). In this case the instantaneous spatial structure is single-well as in region II in fig.~\ref{pic:diagram-r,sigma-chim}. The temporal dynamics, however, is characterized by chaotic switching between the wells, which all the elements perform simultaneously. 
%
%%%%%%%%%%%%%%%%%%%%%%%%%%%% Fig.3 %%%%%%%%%%%%%%%%%%%%%%%%%%%%%%
%
\begin{figure}[!ht]
\centering
\includegraphics[width=0.6\linewidth]{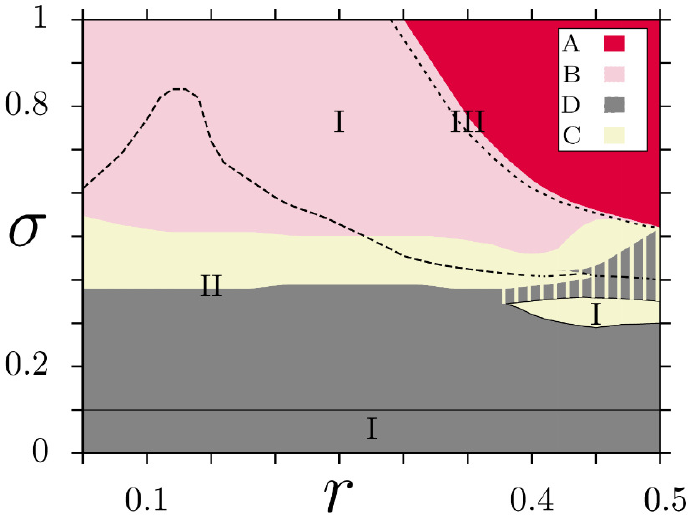}
\caption{Map of regimes in system Eq.\eqref{eq:main} in the plane of parameters $(r,\sigma)$. A: full chaotic synchronization; B: partial coherence; C: chimera states; D: spatial incoherence; I: dominance of double-well spatial structures; II: single-well structures; III: synchronization of chaotic switching. Parameters: $N=100$, $\alpha=3$, $\beta=10$. (Color online)}
\label{pic:diagram-r,sigma-chim}
\end{figure}
%
%%%%%%%%%%%%%%%%%%%%%%%%%%%%%%%%%%%%%%%%%%%%%%%%%%%%%%%%%%%%%%%%%%%%%%%%%%%

As it can be seen from the diagram in fig. \ref{pic:diagram-r,sigma-chim} the bottom border of single-well structures in the region II (solid line in fig.~\ref{pic:diagram-r,sigma-chim}) and the border of the incoherent region D are almost independent of the coupling radius $r$.
Also chimera states can be found for almost all considered values of the coupling range $r$ and intermediate values of the coupling strength $\sigma$ (region C in fig.~\ref{pic:diagram-r,sigma-chim}). Interestingly, for large coupling range $r>0.38$ the region of chimera states becomes larger and chimera patterns begin to alternate with incoherent structures (alternating gray and light-yellow (\textit{online}) strips in the diagram fig.~\ref{pic:diagram-r,sigma-chim}). A top border of the single-well region II (dotted line) has a relatively complex shape and the border of the full chaotic synchronization (region A) shifts dramatically towards large values of coupling strength. For $r<0.3$ the full chaotic synchronization disappears and the partial coherent patterns are observed (region B). When coupling radius is very large ($r\approx 0.5$) the regime of full chaotic synchronization occurs right after the chimera regime while increasing coupling strength $\sigma$. The so-called blowout bifurcation \cite{Ott-1994, Astakhov-1997} takes place at the border between the regions of partial coherence and full chaotic synchronization.
   
The diagram shown in fig.~\ref{pic:diagram-r,sigma-chim} uncovers mainly spatial features of the observed regimes. The temporal dynamics of the lattice elements can be both chaotic and regular. The chaotic behavior in time is typical for the part of the region D located inside the region I while the temporal dynamics in the other parts of region D located in the region II can be both chaotic and regular. This also applies to the regime of chimera states C. The oscillations in the region of partial coherence B are mainly regular (periodic or quasi-periodic) while in the case of full chaotic synchronization A they correspond to the regime of merged chaos demonstrated by every single cubic map. The examples of spatial structures and temporal profiles of the oscillations observed in different regions of the diagram are discussed in the next section. 

For the special type of initial conditions when the values $x_{ij}(t_0)$ for the individual elements of the lattice are given by a slightly randomized harmonic functions regular standing waves
can be observed in the plane of coupling parameters $\sigma$ and $r$. These structures occur in the regime of partial coherence B, have smooth regular instantaneous spatial profile and periodic dynamics in time. The regime of standing waves of different wavelength have been previously reported for the rings of logistic maps and R\"{o}ssler oscillators \cite{Omelchenko-2011, Omelchenko-2012, Semenova-2015}. They are characterized by the wave number $k$ which corresponds to the number of wavelengths that fit into the length of the system. In the case of 2D lattice, each structure is described by two numbers $k_{1}$ and $k_{2}$ which define the number of wavelength in longitudinal and cross sections of the lattice.

Here we restrict our study to the case of $k_{1}=0,~ k_{2}=1$. The results are shown as a two-parametric diagram in fig.~\ref{pic:diagram-r,sigma,timesync} (region $B_1$).
%
%%%%%%%%%%%%%%%%%%%%%%%%%%%%%%%%%%% Fig.4 %%%%%%%%%%%%%%%%%%%%%%%%%%%%%%%%%%%%%%%%%%%%%%%%%
\begin{figure}[!ht]
\centering
\includegraphics[width=0.95\linewidth]{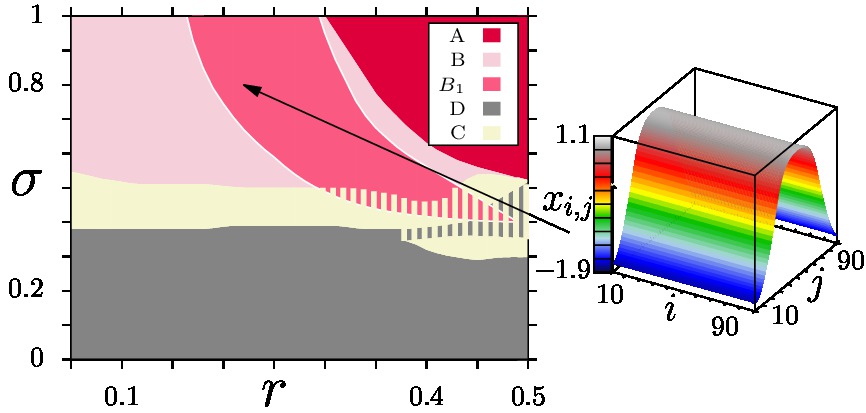}
\caption{Map of regimes in system Eq.\eqref{eq:main} in the plane of parameters $(r,\sigma)$ (see fig.~\ref{pic:diagram-r,sigma-chim}) with the additional area $B_1$, where the standing wave with $k_1=0,~k_2=1$ is realized for specially prepared initial conditions. The snapshot of variable $x_{i,j}$ for the standing wave regime and $r=0.2,~\sigma=0.8$ is shown on the right. Parameters: $N=100$, $\alpha=3$, $\beta=10$. (Color online)}
\label{pic:diagram-r,sigma,timesync}
\end{figure}
%
%%%%%%%%%%%%%%%%%%%%%%%%%%%%%%%%%%%%%%%%%%%%%%%%%%%%%%%%%%%%%%%%%%%%%%%%%%%%%%%%%%%%%%%%%

While moving from top to bottom and right to left we observe period doubling bifurcations. The examples given in fig.~\ref{pic:SeveralSpProf1Di,k=1} illustrate these bifurcations. The instantaneous spatial profiles for the different time moments are represented in fig.~\ref{pic:SeveralSpProf1Di,k=1} for the fixed value of coupling strength and different values of coupling radius $r$. For $r=0.2$ (see fig.~\ref{pic:SeveralSpProf1Di,k=1}(a)) spatial profile does not change in time. This corresponds to the fixed point of the map. For $r=0.19$ (see fig.~\ref{pic:SeveralSpProf1Di,k=1}(b)) there occur two profiles repeating over one iteration. We, therefore, observe periodic oscillations with period $2$. Further decreasing the coupling radius $r=0.18$ (see fig.~\ref{pic:SeveralSpProf1Di,k=1}(c)) we observe many different profiles which signify the appearance of chaotic oscillations.
%
%%%%%%%%%%%%%%%%%%%%%%%%%%%%%%%%%%%% Fig.5 %%%%%%%%%%%%%%%%%%%%%%%%%%%%%%%%%%%%%%%%%%%%%
\begin{figure}[!ht]
\centering
\parbox[c]{.28\linewidth}{
  \includegraphics[width=\linewidth]{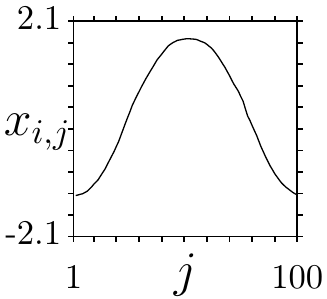}
  \center (a)
}
\parbox[c]{.28\linewidth}{
  \includegraphics[width=\linewidth]{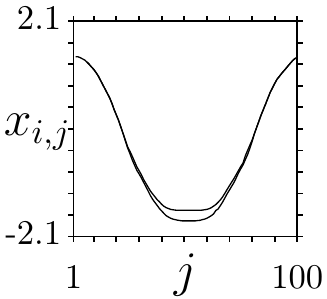}
  \center (b)  
}
\parbox[c]{.28\linewidth}{
  \includegraphics[width=\linewidth]{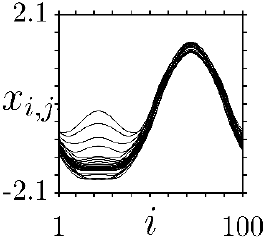}
  \center (c)
}
\caption{A set of 30 instantaneous spatial cutoffs for the variable $x_n(j)$ at different time moments in the regime of standing wave for fixed $i=50$, $k_{1}=0,~ k_{2}=1$, $\sigma=0.8$ and different values of coupling radius (a): $r=0.2$, (b): $r=0.19$, (c): $r=0.18$. Parameters:  $N=100$, $\alpha=3$, $\beta=10$.}
\label{pic:SeveralSpProf1Di,k=1}
\end{figure}
%
%%%%%%%%%%%%%%%%%%%%%%%%%%%%%%%%%%%%%%%%%%%%%%%%%%%%%%%%%%%%%%%%%%%%%%%%%%%%%%%%%%%%%%%%%

Besides the considered structure $k_{1}=0,~k_{2}=1$ in the lattice Eq.~\eqref{eq:main} one can obtain many other similar patterns which are observed in more narrow regions in the plane of parameters. Instantaneous spatial profile cutoffs demonstrating a selection of such structure are shown in fig.~\ref{pic:SeveralSpProf1Di,dif_k}.
%
%%%%%%%%%%%%%%%%%%%%%%%%%%%%%%%%%% Fig.6 %%%%%%%%%%%%%%%%%%%%%%%%%%%%%%%%%%%%
%
\begin{figure}[!ht]
\centering
\parbox[c]{.28\linewidth}{
  \includegraphics[width=\linewidth]{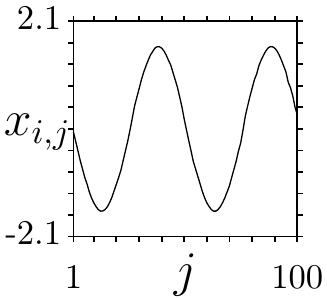}
  \center (a)
}
\parbox[c]{.28\linewidth}{
  \includegraphics[width=\linewidth]{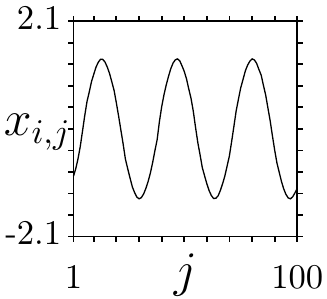}
  \center (b)
}
\parbox[c]{.28\linewidth}{
  \includegraphics[width=\linewidth]{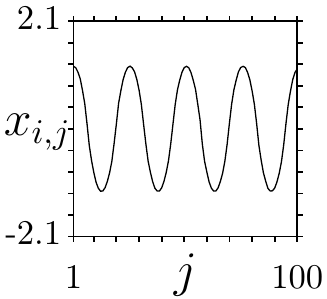}
  \center (c)
}
\caption{Instantaneous spatial cutoffs for the variable $x_n(j)$ in the regime of standing wave for fixed $i=50$, $k_{1}=0,~ k_{2}=2,3,4$, $\sigma=0.8$ and different values of coupling radius (a): $r=0.11$, (b): $r=0.08$, (c): $r=0.07$. Parameters:  $N=100$, $\alpha=3$, $\beta=10$.}
\label{pic:SeveralSpProf1Di,dif_k}
\end{figure}
%%%%%%%%%%%%%%%%%%%%%%%%%%%%%%%%%%%%%%%%%%%%%%%%%%%%%%%%%%%%%%%%%%%%%%%%%%%%%%%%%%

\section{Dynamic regimes for different values of coupling strength $\sigma$ and fixed coupling radius $r=0.35$}

Next we fix the coupling radius $r=0.35$ since this particular value allows to observe the most typical regimes of the system by varying coupling strength $\sigma$. Further we consider the behavior of the system for increasing coupling strength $\sigma$ from $0$ to $1$ in more detail.

For this purpose we calculate phase-parametric diagram for one selected element of the lattice $i=j=50$ (fig.~\ref{pic:PhaseDiagram-x,r=0_35}) by varying parameter $\sigma$ for the fixed $r$. 
As initial condition for each value of the control parameter we take the same realization of the random sequence in the interval $[0,1]$ (similarly to the two-parametric diagram in fig. \ref{pic:diagram-r,sigma-chim}. For the fixed value of parameter $\sigma$ we obtain the values $x_{i,j},~ i=50, ~j=50$ at different time moments (dots in fig.~\ref{pic:PhaseDiagram-x,r=0_35}). 
It is important to note that such a diagram depends on the selected element and initial conditions. However, for any element of the lattice one obtains qualitatively the same diagram as for the node $i=j=50$. The phase-parametric diagram allows, therefore, to estimate the borders between various spatial structures. 
%
%%%%%%%%%%%%%%%%%%%%%%%%%%%%%%%%% Fig.7 %%%%%%%%%%%%%%%%%%%%%%%%%%%%%%%%%%%%%%%%%%%%%%%%
\begin{figure}[!ht]
\centering
\includegraphics[width=1.0\linewidth]{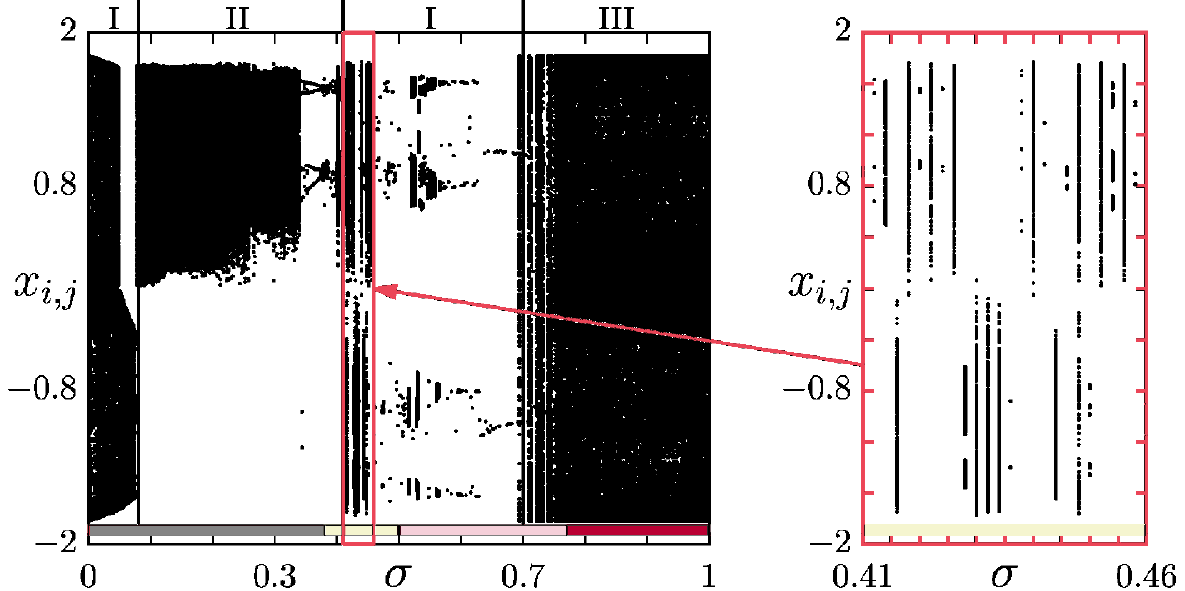}
\caption{
Phase-parametric diagram of the lattice element $i=50,~j=50$ for varying coupling strength $\sigma$ and fixed $r=0.35$. The regions indicated by roman numbers (top) and color code (bottom) correspond to the same regimes in fig.~\ref{pic:diagram-r,sigma-chim}. The panel on the right shows a zoomed-in fragment of the diagram. Parameters:  $N=100$, $\alpha=3$, $\beta=10$.}
\label{pic:PhaseDiagram-x,r=0_35}
\end{figure}
%
%%%%%%%%%%%%%%%%%%%%%%%%%%%%%%%%%%%%%%%%%%%%%%%%%%%%%%%%%%%%%%%%%%%%%%%%%%%%%%%%%%%%%%

\begin{figure}[!ht]
\centering
\parbox[c]{.55\linewidth}{
  \includegraphics[width=\linewidth]{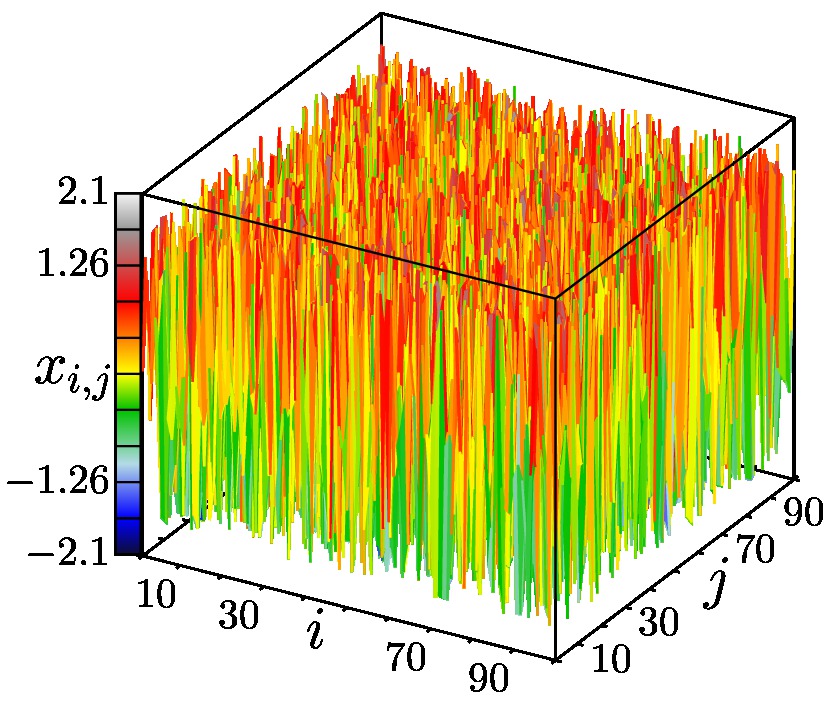}
  \center (\small a)
}
\parbox[c]{.25\linewidth}{
  \includegraphics[width=\linewidth]{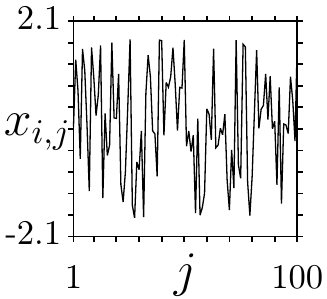}
  \center (\small b)
  \includegraphics[width=\linewidth]{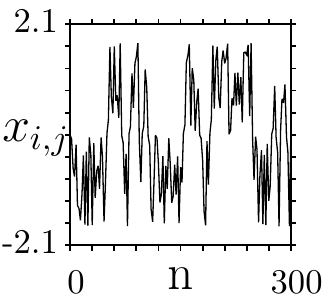}
  \center (\small c)
}
\caption{Regime of spatially incoherent merged chaos for $\sigma=0.044$. (a): 3D snapshot of the variable $x_{i,j}$, (b): instantaneous spatial cutoff $x_{i,j}(j)$ for fixed $i=50$, (c): time realization for one selected element $i=50,~ j=50$. Parameters:  $r=0.35$, $N=100$, $\alpha=3$, $\beta=10$.}
\label{P35_sig0044_comb_chaos}
\end{figure}

For varying coupling strength one observes regimes, corresponding to double-well and single-well structures as well as the regime of synchronization of chaotic switching between positive and negative wells shown as regions I, II, III in fig.~\ref{pic:PhaseDiagram-x,r=0_35}, respectively (for the convenience we use here the same notations and color code (color \textit{online}) as in fig.~\ref{pic:diagram-r,sigma-chim}). In the case of weak coupling we detect double-well spatial structure (region I). However, the behavior of the lattice elements in time can be different. For the small values of coupling strength ($\sigma < 0.05$) the dynamics of each element corresponds to the regime of merged chaos and, therefore, $x_{i,j}$ can be both positive and negative for different iteration. This behavior is observed not only for the previously described type of initial conditions, but also for any other choice of initial conditions for the lattice elements. The illustration of this regime is provided by fig.~\ref{P35_sig0044_comb_chaos}.
%
%%%%%%%%%%%%%%%%%%%%%%%%%%%%%%%%%%% Fig.8 %%%%%%%%%%%%%%%%%%%%%%%%%%%%%%%%%%%%%
%
%
%%%%%%%%%%%%%%%%%%%%%%%%%%%%%%%%%%%%%%%%%%%%%%%%%%%%%%%%%%%%%%%%%%%%%%%%%%%%%%%%%

\begin{figure}[!ht]
\centering
\parbox[c]{.55\linewidth}{
  \includegraphics[width=\linewidth]{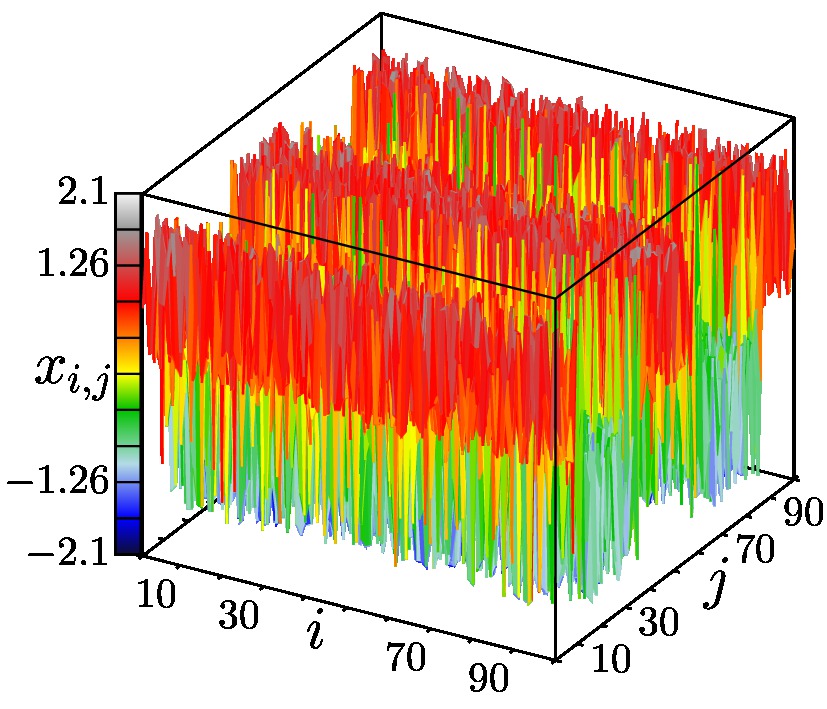}
  \center (\small a)
}
\parbox[c]{.25\linewidth}{
  \includegraphics[width=\linewidth]{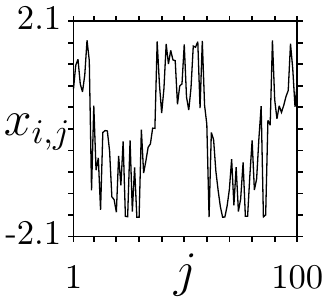}
  \center (\small b)
  \includegraphics[width=\linewidth]{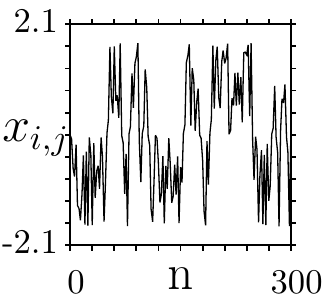}
  \center (\small c)
}
\caption{Double-well chaotic structure with bistable temporal dynamics of the elements for $\sigma=0.048$. (a): 3D snapshot the of variable $x_{i,j}$, (b): instantaneous spatial cutoff $x_{i,j}(j)$ for fixed $i=50$, (c): time realization $x_{i,j}(n)$ for one selected element $i=50,~ j=50$. Parameters:  $r=0.35$, $N=100$, $\alpha=3$, $\beta=10$.}
\label{P35_sig0048_comb_to_bist}
\end{figure}

The temporal dynamics of the lattice elements becomes bistable as the coupling strength reaches the value $\sigma\approx 0.05$. Each element remains in either positive or negative well (positive or negative states). At the same time, the double-well structures are observed for arbitrary chosen initial conditions: one group of elements oscillates in the positive well while the other group performs oscillations in the negative well (fig.~\ref{P35_sig0048_comb_to_bist}). As shown in fig.~\ref{pic:PhaseDiagram-x,r=0_35} the element $i=50,~j=50$ is located in the negative (left) well for $0.05 < \sigma < 0.095$. However, it switches to the positive (right) well for $\sigma\approx 0.095$. The temporal dynamics remains chaotic and is characterized by a rather uniform distribution of the values $x_{i,j}$ between $0$ and $1.8$.
%
%%%%%%%%%%%%%%%%%%%%%%%%%%%%%% Fig.9 %%%%%%%%%%%%%%%%%%%%%%%%%%%%%%%%%%%%%%%%%%%%%%%%
%
%
%%%%%%%%%%%%%%%%%%%%%%%%%%%%%%%%%%%%%%%%%%%%%%%%%%%%%%%%%%%%%%%%%%%%%%%%%%%%%%%%%%%%%%%%

\begin{figure}[!ht]
\centering
\parbox[c]{.55\linewidth}{
  \includegraphics[width=\linewidth]{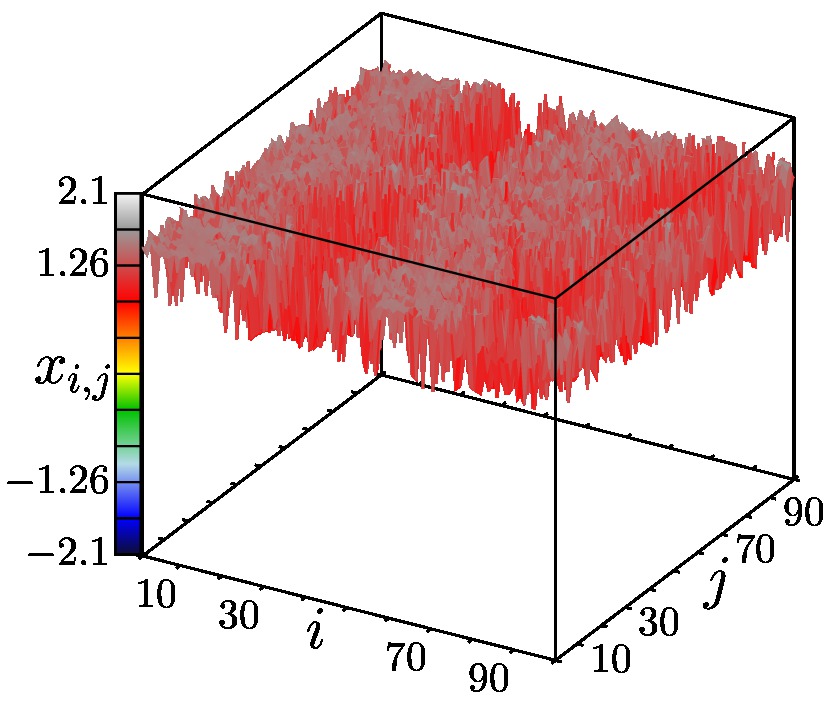}
  \center (\small a)
}
\parbox[c]{.25\linewidth}{
  \includegraphics[width=\linewidth]{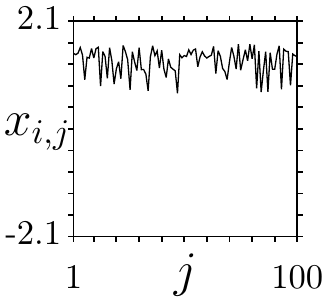}
  \center (\small b)
  \includegraphics[width=\linewidth]{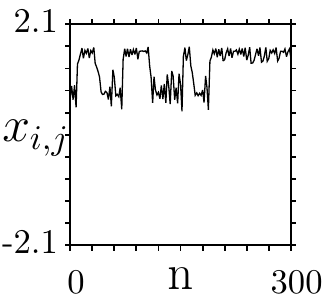}
  \center (\small c)
}
\caption{Single-well structure in the regime of bistable chaos for $\sigma=0.284$. (a): 3D snapshot of the variable $x_{i,j}$, (b) instantaneous spatial cutoff $x_{i,j}(j)$ for fixed $i=20$, (c) time realization $x_{i,j}(n)$ for one selected element $i=50,~ j=50$. Parameters:  $r=0.35$, $N=100$, $\alpha=3$, $\beta=10$.}
\label{P35_sig0284_top_chaos}
\end{figure}

As coupling strength reaches the value $\sigma \approx 0.1$ there occurs a transition to the regime of single-well structures (region II in the diagram). Here we observe the appearance of the complex but stable in time structures characterized by chaotic single-well dynamics of the elements both in time and in space. In other words all the lattice elements are localized either in negative or positive well where they oscillate during all the observation time (fig.~\ref{P35_sig0284_top_chaos}). Such behavior is observed in the bottom part of the region II (single-well structures) corresponding to the regime of incoherence D (dark-gray region in fig.~\ref{pic:diagram-r,sigma-chim}.  
%
%%%%%%%%%%%%%%%%%%%%%%%%%%%%%% Fig.10 %%%%%%%%%%%%%%%%%%%%%%%%%%%%%%%%%%%%%%%%%%%%%%%%
%
%
%%%%%%%%%%%%%%%%%%%%%%%%%%%%%%%%%%%%%%%%%%%%%%%%%%%%%%%%%%%%%%%%%%%%%%%%%%%%%%%%%%%%%%%%

For further increasing coupling strength the correlations between the elements increase resulting in the formation of clusters with qualitatively different behavior of neighboring elements. In particular, one part of the lattice elements is characterized by incoherent behavior of the neighboring nodes while the other part demonstrates synchronization. The appearance of stationary in time coherent and incoherent clusters indicates the transition of the system to the regime of chimera state (region C (light-yellow color) in fig.~\ref{pic:diagram-r,sigma-chim}).
 
\begin{figure}[!ht]
\centering
\parbox[c]{.49\linewidth}{
  \includegraphics[width=\linewidth]{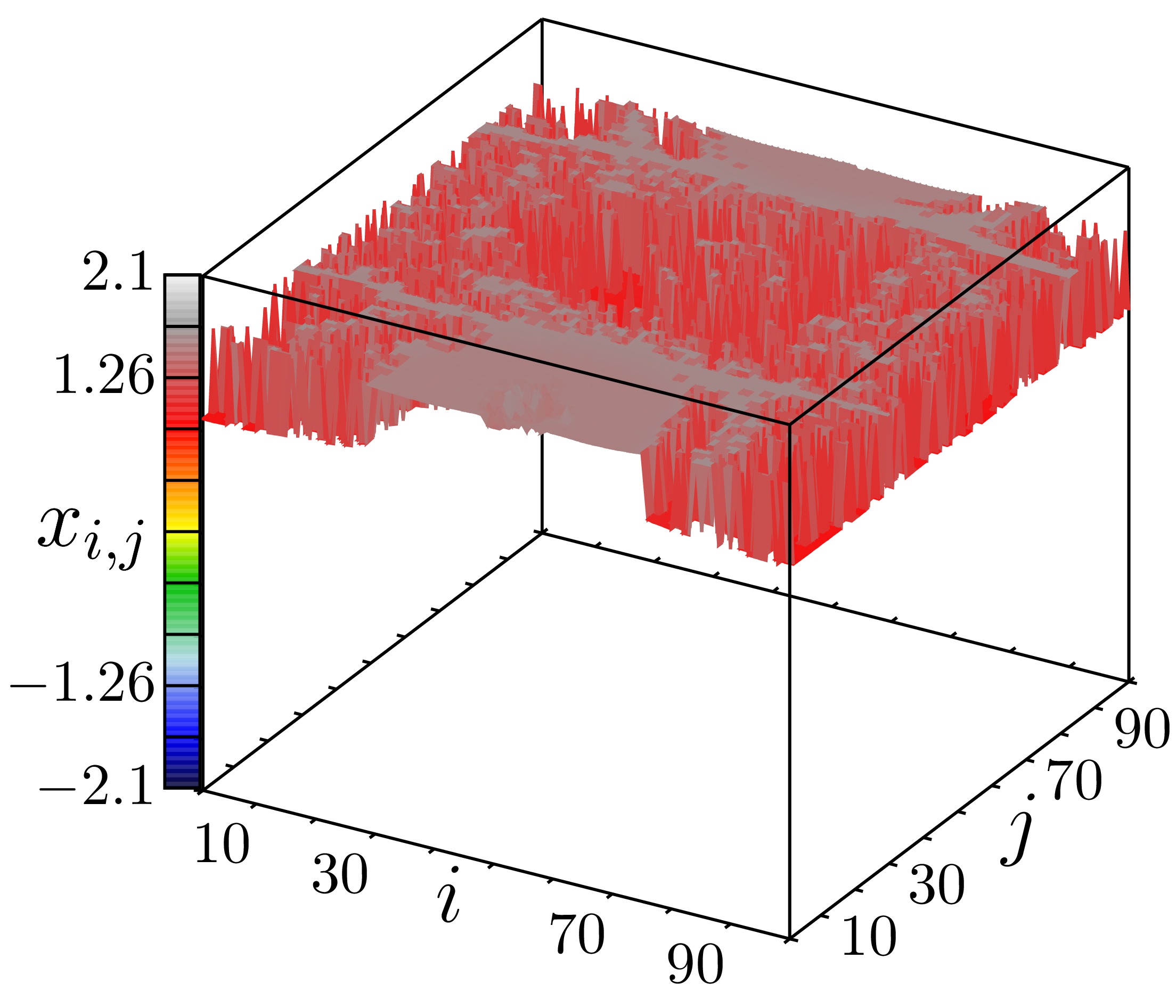}
  \center (\small a)
}
\parbox[c]{.23\linewidth}{
  \includegraphics[width=\linewidth]{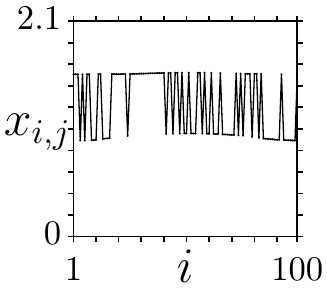}
  \center (\small b)
  \includegraphics[width=\linewidth]{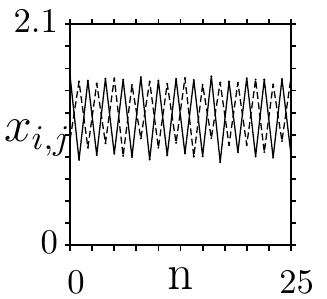}
  \center (\small c)
}
\parbox[c]{.23\linewidth}{
  \includegraphics[width=\linewidth]{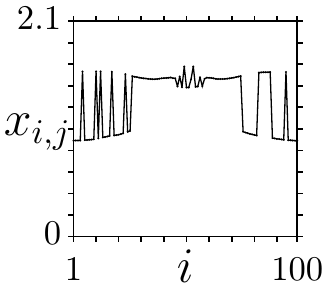}
  \center (\small d)
  \includegraphics[width=\linewidth]{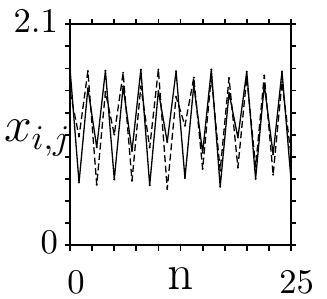}
  \center (\small e)
}
\caption{Single-well structure in the regime of phase and amplitude chimeras for $\sigma=0.402$. (a): 3D snapshot of the variable $x_{i,j}$, (b): instantaneous spatial cutoff $x_{i,j}(i)$ for fixed $j=80$, (c): time realization $x_{i,j}(n)$ for two selected neighboring elements  $i=51,~ j=80$ (solid line) and $i=51,~ j=81$ (dashed line) belonging to the incoherent domain of the phase chimera, (c): instantaneous spatial cutoff $x_{i,j}(i)$ for fixed $j=6$, (d) time realization $x_{i,j}(n)$ for two selected neighboring elements $i=49,~j=6$ (solid line) and $i=50,~ j=6$ (dashed line) belonging to the incoherent domain of the amplitude chimera. Parameters:  $r=0.35$, $N=100$, $\alpha=3$, $\beta=10$.}
\label{P35_sig0402_two_chimeras}
\end{figure}

\begin{figure}[!ht]
\centering
\parbox[c]{.49\linewidth}{
  \includegraphics[width=\linewidth]{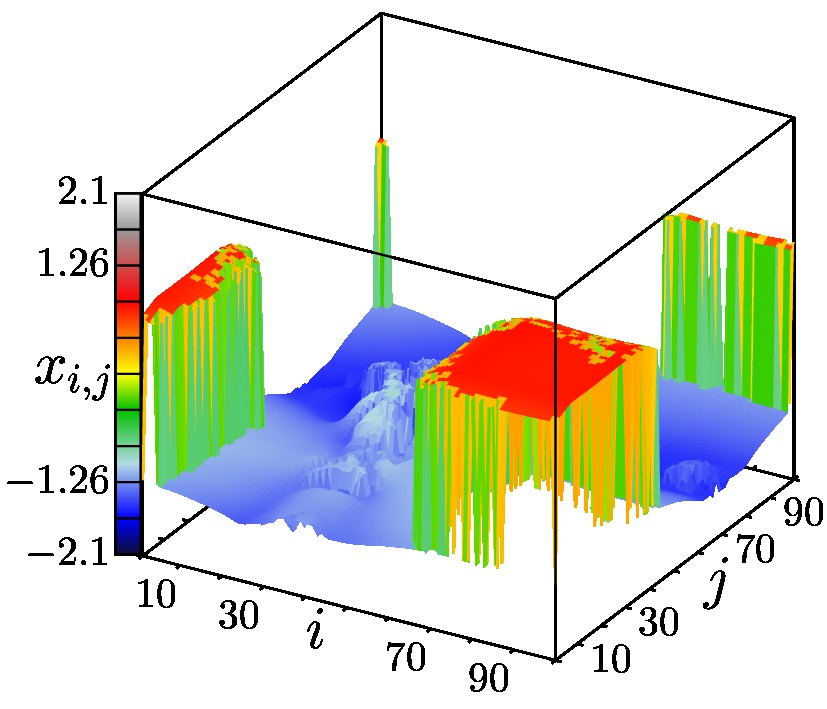}
  \center (\small a)
}
\parbox[c]{.23\linewidth}{
  \includegraphics[width=\linewidth]{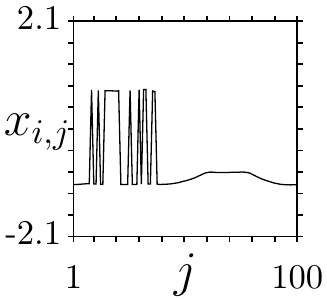}
  \center (\small b)
  \includegraphics[width=\linewidth]{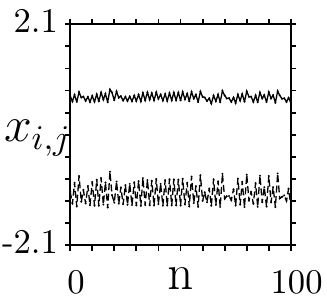}
  \center (\small c)
}
\parbox[c]{.23\linewidth}{
  \includegraphics[width=\linewidth]{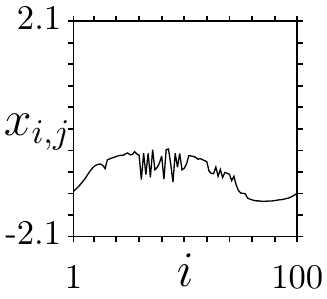}
  \center (\small d)
  \includegraphics[width=\linewidth]{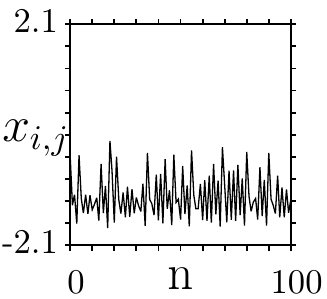}
  \center (\small e)
}

\caption{Double-well structure in the regime of double-well chimeras and amplitude chimeras for $\sigma=0.444$. (a): 3D snapshot  of the variable $x_{i,j}$, (b): instantaneous spatial cutoff $x_{i,j}(j)$ for fixed $i=6$, (c): time realization $x_{i,j}(n)$ for two selected neighboring elements $i=49,~ j=6$ (solid line) and $i=50,~j=6$ (dashed line) belonging to the incoherent domain of double-well chimera, (d): instantaneous spatial cutoff $x_{i,j}(i)$ for fixed $j=50$, (e) time realization $x_{i,j}(n)$ for one selected element $i=31,~ j=50$ belong to the incoherent domain of amplitude chimera. Parameters:  $r=0.35$, $N=100$, $\alpha=3$, $\beta=10$.}
\label{P35_sig0444_comb_chim}
\end{figure}

The regime where regions II and C overlap corresponds to chimera states which are characterized by single-well dynamics in space and time. It means that all the lattice elements are located either in positive or negative well. These chimera states which we find here for the lattice are similar to the phase and amplitude chimeras previously detected for the rings of logistic maps and chaotic self-sustained oscillators \cite{Bogomolov-2016,Anishchenko-2016,Anishchenko-2016b}. The illustrative example of the structure involving phase and amplitude chimeras is shown in fig.~\ref{P35_sig0402_two_chimeras}. In the case of phase chimera the elements from the incoherent domains oscillate with the shift of one iteration which corresponds to the to the shift of one half of a period for a continuous-time system. The oscillations are ``in phase'' for amplitude chimera but their ``amplitudes'' (instantaneous values $x_{i,j}$) are very different.
%
%%%%%%%%%%%%%%%%%%%%%%%%%%%%%% Рис.11 %%%%%%%%%%%%%%%%%%%%%%%%%%%%%%%%%%%%%%%%%%%%%%%%
%
%
%%%%%%%%%%%%%%%%%%%%%%%%%%%%%%%%%%%%%%%%%%%%%%%%%%%%%%%%%%%%%%%%%%%%%%%%%%%%%%%%%%%%%%%%
%AZ READ THIS PART ONE MORE TIME
On the top border of the single-well region (dashed line in the diagrams) there occurs a bifurcation resulting in the merging of the chimera states belonging to different wells and
a new type of chimera pattern typical for the intersection of the regions I and C arises. In this pattern the elements of coherent domains are localized in one well while the elements of the incoherent clusters are distributed incoherently between the wells (fig.~\ref{P35_sig0444_comb_chim}). Therefore, we call this pattern \textit{double-well chimera}. Such chimera states can not be obtained in ensembles of logistic maps or chaotic systems of the R\"{o}ssler type. The instantaneous spatial cutoff $x_{i,j}(j)$ for the fixed $i=6$ provides a clear representation of the double-well chimera regime. Time realizations which are given in fig.~\ref{P35_sig0444_comb_chim}(c) for two neighboring elements from the incoherent cluster show that the elements are localized in different wells during the whole observation time. The double-well chimera coexists with the amplitude chimera located in the negative well. It can be clearly seen in the instantaneous spatial cutoff for $i=50$. It is important to note that chimera structures observed in the lattice Eq. (\ref{eq:main}) mainly combine clusters of different chimera types: phase, amplitude and double-well chimeras. Various spatial distributions of coherent and incoherent clusters can be obtained in different parts of a 2D spatial profile.

In the regime of double-well chimera the element of the lattice $i=50,~j=50$ visits both wells, in spite of the initial conditions chosen in the positive well for all the elements (right panel in fig.~\ref{pic:PhaseDiagram-x,r=0_35}). Moreover, as it can be seen from the figure the considered element can switch to another well even for the small shift of parameter. Therefore, spatial structures in the regime of double-well chimera are very sensitive to both initial conditions and variations of the coupling strength. Once an element of the lattice reaches a particular well it remains there during the whole observation time.  
%
%%%%%%%%%%%%%%%%%%%%%%%%%%%%%% Fig.12 %%%%%%%%%%%%%%%%%%%%%%%%%%%%%%%%%%%%%%%%%%%%%%%%
%
%
%%%%%%%%%%%%%%%%%%%%%%%%%%%%%%%%%%%%%%%%%%%%%%%%%%%%%%%%%%%%%%%%%%%%%%%%%%%%%%%%%%%%%%%%%

\begin{figure}[!ht]
\centering
\parbox[c]{.55\linewidth}{
  \includegraphics[width=\linewidth]{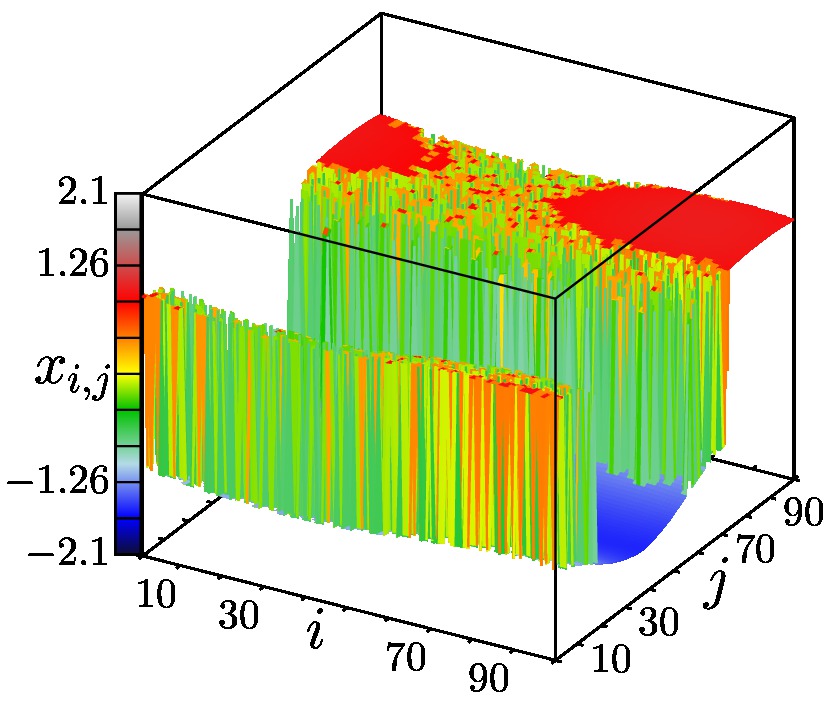}
  \center (\small a)
}
\parbox[c]{.25\linewidth}{
  \includegraphics[width=\linewidth]{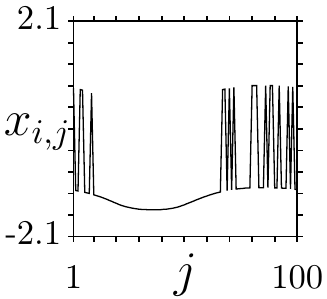}
  \center (\small b)
  \includegraphics[width=\linewidth]{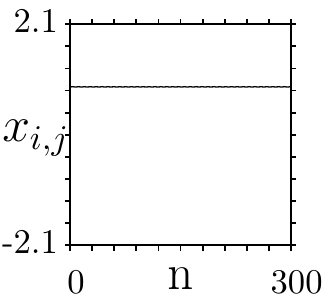}
  \center (\small c)
}
\caption{Stationary double-well chimera for $\sigma=0.450$. (a): 3D snapshot of the variable $x_{i,j}$, (b): instantaneous spatial cutoff $x_{i,j}(j)$ for fixed $i=32$, (c): time realization $x_{i,j}(n)$ for one selected element $i=50,~ j=50$. Parameters:  $r=0.35$, $N=100$, $\alpha=3$, $\beta=10$.}
\label{P35_sig0450_comb_chim}
\end{figure}

If we slightly increase the coupling parameter there occur stationary spatial structures which are similar to those observed in ensembles of harmonic self-sustained oscillators in the regime of chimera death (or incoherent oscillation death \cite{SCH15b}). For these structures we do not observe any temporal dynamics (fig.~\ref{P35_sig0450_comb_chim}(c)).
However, there is coexistence in space of coherent domains where the elements are localized in one and the same well and incoherent domains where neighboring elements are randomly distributed between the two wells. therefore, we call these patterns \textit{double-well chimera death}.

%
%%%%%%%%%%%%%%%%%%%%%%%%%%%%%%%%%%%% Fig.13 %%%%%%%%%%%%%%%%%%%%%%%%%%%%%%%%%%%%%%%
%
%
%%%%%%%%%%%%%%%%%%%%%%%%%%%%%%%%%%%%%%%%%%%%%%%%%%%%%%%%%%%%%%%%%%%%%%%%%%%%%%%%%%%

\begin{figure}[!ht]
\centering
\parbox[c]{.55\linewidth}{
  \includegraphics[width=\linewidth]{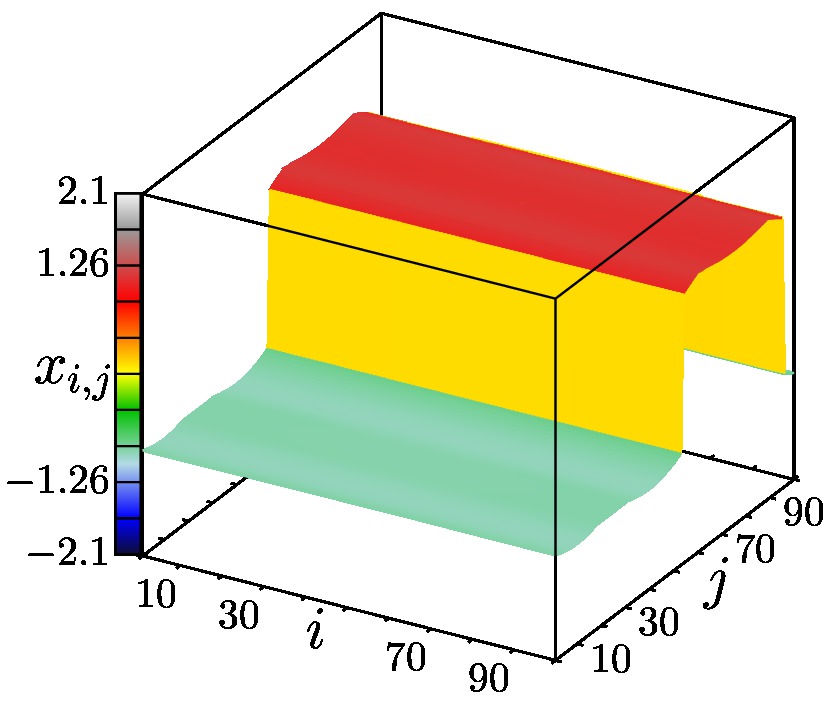}
  \center (\small a)
}
\parbox[c]{.25\linewidth}{
  \includegraphics[width=\linewidth]{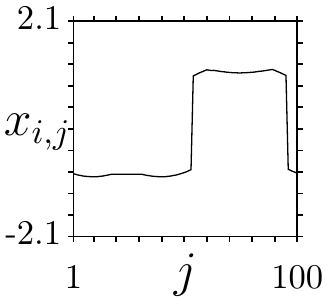}
  \center (\small b)
  \includegraphics[width=\linewidth]{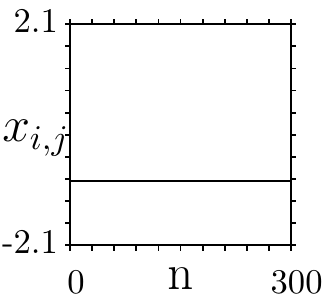}
  \center (\small c)
}
\caption{Stationary double-well spatial structure in the regime of partial coherence for $\sigma=0.502$. (a): 3D snapshot of the variable $x_{i,j}$, (b): instantaneous spatial cutoff $x_{i,j}(j)$ for fixed $i=50$, (c): time realization $x_{i,j}(n)$ for one selected element $i=50,~ j=50$. Parameters:  $r=0.35$, $N=100$, $\alpha=3$, $\beta=10$.}
\label{P35_sig0502_part_sync_comb}
\end{figure}

Once the value $\sigma\approx 0.51$ is reached there occurs a transition to a regime of partial coherence (region B in two-parametric diagram in fig.~\ref{pic:diagram-r,sigma-chim}). The double-well spatial structures with smooth spatial profiles are observed in the region B. Lattice elements are distributed between different wells. The dynamics in time is bistable (single-well) as in the region C. Every element remains in its well and performs regular oscillations corresponding to cycles of different periods. Stationary in time spatial structures of this type also appear with the elements being distributed between two fixed points (fig.~\ref{P35_sig0502_part_sync_comb}). This structure corresponds to a standing wave along the $i$ axis and has a form of a single impulse with sharp transitions between the states belonging two different wells.
%
%%%%%%%%%%%%%%%%%%%%%%%%%%%%%%%%%%%%% Fig.14 %%%%%%%%%%%%%%%%%%%%%%%%%%%%%%%%%%%%%%%%%%%%%
%
%%%%%%%%%%%%%%%%%%%%%%%%%%%%%%%%%%%%%%%%%%%%%%%%%%%%%%%%%%%%%%%%%%%%%%%%%%%%%%%%%%%%%%%

\begin{figure}[!ht]
\centering
\parbox[c]{.55\linewidth}{
  \includegraphics[width=\linewidth]{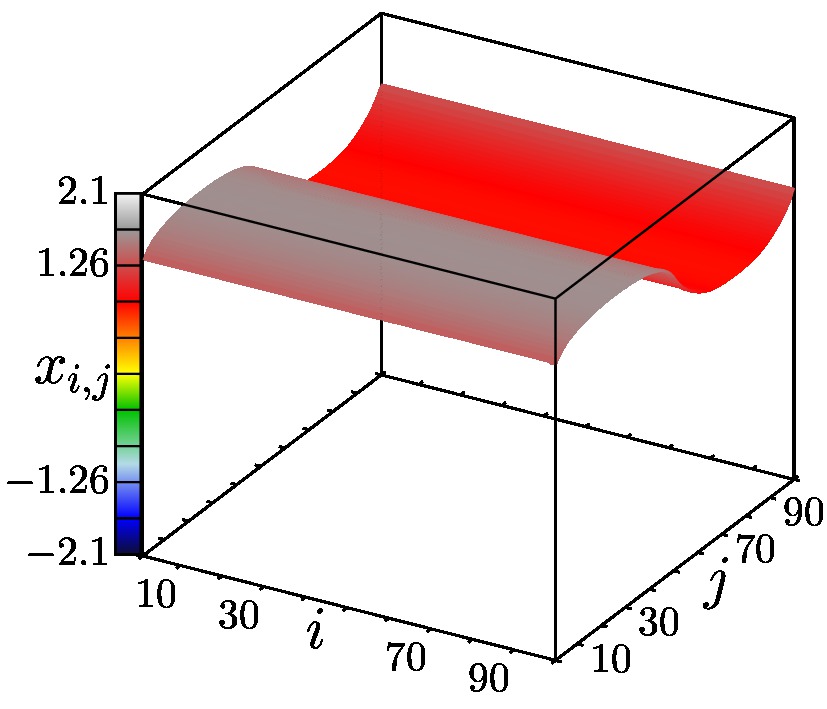}
  \center (\small a)
}
\parbox[c]{.25\linewidth}{
  \includegraphics[width=\linewidth]{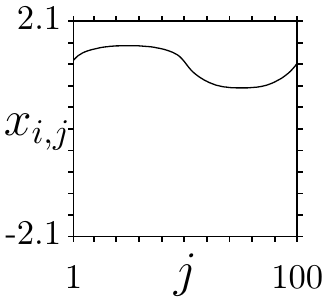}
  \center (\small b)
  \includegraphics[width=\linewidth]{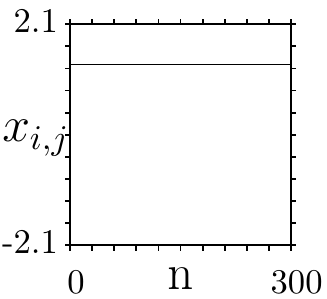}
  \center (\small c)
}
\caption{Stationary single-well spatial structure for $\sigma=0.614$. (a): 3D snapshot of the variable $x_{i,j}$, (b): instantaneous spatial cutoff $x_{i,j}(j)$ for fixed $i=50$, (c): time realization $x_{i,j}(n)$ for one selected element $i=50,~ j=50$. Parameters:  $r=0.35$, $N=100$, $\alpha=3$, $\beta=10$.}
\label{P35_sig0614_part_sync_top}
\end{figure}

Although in the region I the initial conditions randomly distributed within one well favor double-well structures, for some cases one obtains stationary structures with the elements localized in one well (fig.~\ref{P35_sig0614_part_sync_top}).
%
%%%%%%%%%%%%%%%%%%%%%%%%%%%%%%%%% Fig.15 %%%%%%%%%%%%%%%%%%%%%%%%%%%%%%%%%%%%%%%
%
%
%%%%%%%%%%%%%%%%%%%%%%%%%%%%%%%%%%%%%%%%%%%%%%%%%%%%%%%%%%%%%%%%%%%%%%%%%%%%%%%%%%%%%%%

\begin{figure}[!ht]
\centering
\parbox[c]{.55\linewidth}{
  \includegraphics[width=\linewidth]{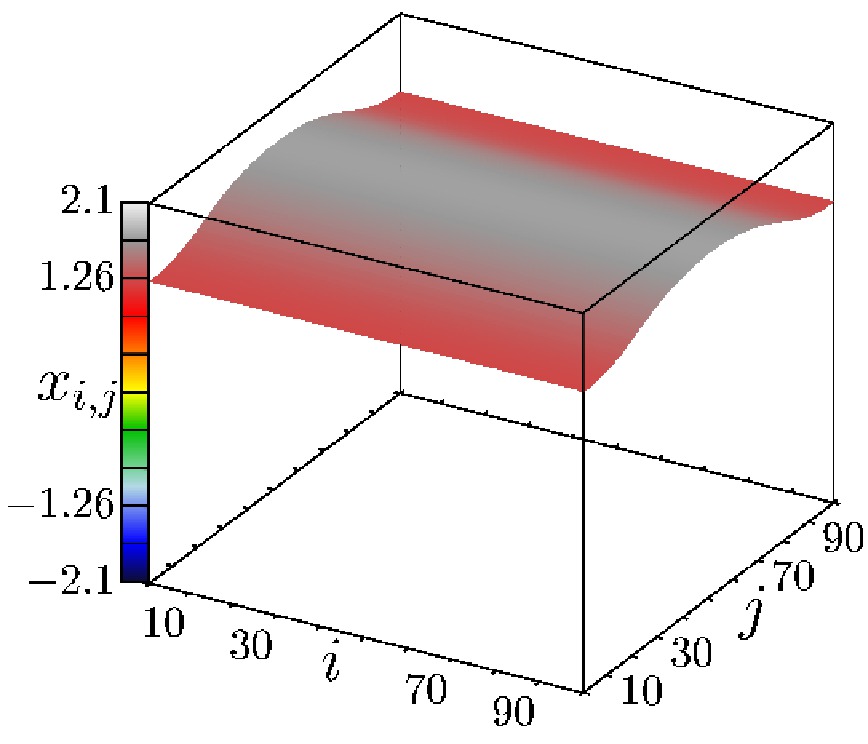}
  \center (\small a)
}
\parbox[c]{.25\linewidth}{
  \includegraphics[width=\linewidth]{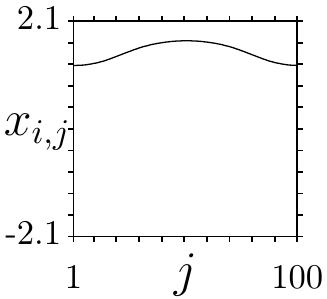}
  \center (\small b)
  \includegraphics[width=\linewidth]{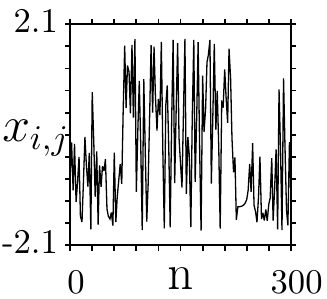}
  \center (\small c)
}
\caption{Synchronization of chaotic switching for $\sigma=0.714$. (a): 3D snapshot of the variable $x_{i,j}$, (b): instantaneous spatial cutoff $x_{i,j}(j)$ for fixed $i=50$, (c): time realization $x_{i,j}(n)$ for one selected element  $i=50,~ j=50$. Parameters:  $r=0.35$, $N=100$, $\alpha=3$, $\beta=10$.}
\label{P35_sig0714_part_sync}
\end{figure}

While crossing the border of the region III (dashed line in the diagram shown in fig.~\ref{pic:diagram-r,sigma-chim}) the mutual correlation of the elements significantly grows and the switching between positive and negative wells occurs simultaneously for all the elements of the lattice. The oscillations in time in this case correspond to the regime of the merged chaos. However, the instantaneous spatial profiles show simultaneous localization of all the elements in one and the same well (fig.~\ref{P35_sig0714_part_sync}). In the regime where the regions III and B overlap the full chaotic synchronization of oscillations is still not reached and the oscillations of different elements are, therefore, not identical.
%
%%%%%%%%%%%%%%%%%%%%%%%%%%%%%%%%%%%% Fig.16 %%%%%%%%%%%%%%%%%%%%%%%%%%%%%%%%%%%%%%%
%
%
%%%%%%%%%%%%%%%%%%%%%%%%%%%%%%%%%%%%%%%%%%%%%%%%%%%%%%%%%%%%%%%%%%%%%%%%%%%%%%%%%%%%

\begin{figure}[!ht]
\centering
\parbox[c]{.55\linewidth}{
  \includegraphics[width=\linewidth]{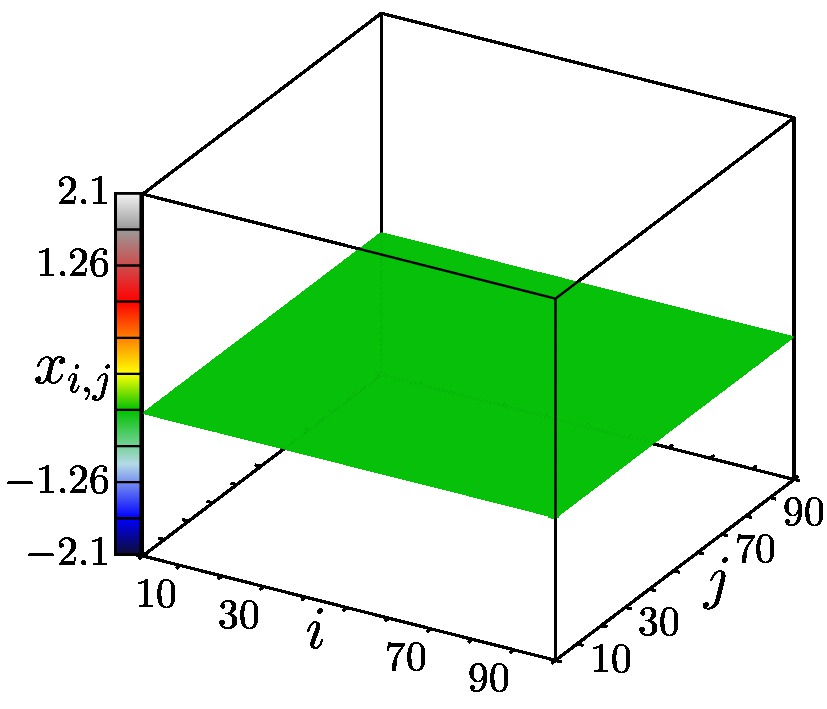}
  \center (\small a)
}
\parbox[c]{.25\linewidth}{
  \includegraphics[width=\linewidth]{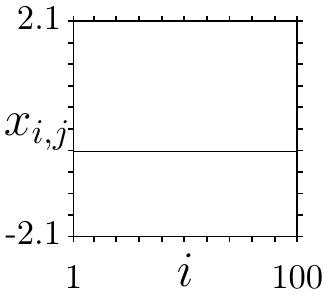}
  \center (\small b)
  \includegraphics[width=\linewidth]{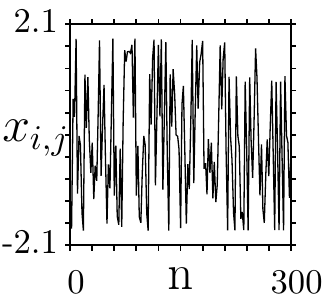}
  \center (\small c)
}
\caption{Full chaotic synchronization for $\sigma=0.894$. (a): 3D snapshot of the variable $x_{i,j}$, (b): instantaneous spatial cutoff $x_{i,j}(j)$ for fixed $i=50$, (c): time realization $x_{i,j}(n)$ for one selected element $i=50,~ j=50$. Parameters:  $r=0.35$, $N=100$, $\alpha=3$, $\beta=10$.}
\label{P35_sig0894_full_sync}
\end{figure}

For $\sigma\approx 0.72$ the blowout bifurcation occurs (in the backward direction) and the system Eq. (\ref{eq:main}) reaches the regime of full chaotic synchronization of the merged chaos (region A and fig.~\ref{P35_sig0894_full_sync}).

%
%%%%%%%%%%%%%%%%%%%%%%%%%%%%%%%%%%%% Fig.17 %%%%%%%%%%%%%%%%%%%%%%%%%%%%%%%%%%%%%%%
%
%
%%%%%%%%%%%%%%%%%%%%%%%%%%%%%%%%%%%%%%%%%%%%%%%%%%%%%%%%%%%%%%%%%%%%%%%%%%%%%%%%%%%%%%%

\section*{Conclusion}

In conclusion, we have studied spatio-temporal behavior of a 2D lattice of nonlocally coupled chaotic maps. The distinctive feature of this 2D ensemble is its local dynamics represented by a one-dimensional cubic map (\ref{eq:main}) demonstrating chaotic behavior (see fig.~\ref{pic:diagram-alpha}). In particular, depending on the control parameter the chaotic dynamics of a single element can manifest itself through the regime of bistability or the regime of merging of chaotic attractors (see fig.\ref{pic:diagram-alpha}). For the 2D ensemble we have extensively analyzed the transition from incoherence (spatio-temporal chaos) to coherence (full chaotic synchronization) while increasing the coupling strength $0 \leq \sigma \leq 1$ (fig.~\ref{pic:PhaseDiagram-x,r=0_35}). The boundary conditions are periodic in both directions, initial conditions are randomly distributed within the interval [0,1] which is determined by the attractor size of a single element.

We show that coherence -- incoherence transition in the ensemble (\ref{eq:main}) demonstrates a variety of spatio-temporal structures which can be regular or chaotic with respect to both time and space. For the intermediate range of the coupling range $0.35 < \sigma < 0.55$ and $\alpha = 3.0$ we find chimera states previously detected only in 1D ensembles of chaotic systems and named amplitude and phase chimeras in \cite{Bogomolov-2016,Anishchenko-2016,Anishchenko-2016b}. Moreover, we uncover a novel type of chimera patterns which we call \textit{double-well chimera}. The bistability of the local dynamics in the network plays crucial role for the appearance of double-well chimera state. The interaction of bistable elements in the case of nonlocal coupling leads to the appearance of complex chimera structures characterized by the instantaneous distribution of amplitudes over two regions constituting the bistability regime: positive and negative wells. In more detail, the elements belonging to coherent (synchronous) domains are localized in one well while the nodes from incoherent (asynchronous) clusters switch randomly between the wells. Moreover, we detect a pattern characterized by chimera behaviour of steady sates which we call \textit{double-well chimera death}. The main distinctive feature of chimera patterns we find in 2D ensembles is their combined structure:
different cutoffs of the pattern may demonstrate distinct chimera types: amplitude, phase and double-well chimeras. And even within one cutoff one may detect clusters belonging to distinct types of chimera states. 

The system under study can be characterized by an inhomogeneous spatial distribution of clusters with different structures (see fig.~\ref{P35_sig0444_comb_chim}). This property indicates a typical feature of the cluster distributions with various patterns observed in large ensembles in nature and engineering. Chimera states occurring due to the bistability of local dynamics for nonlocal coupling topologies can be applied to the modeling of the processes in neurodynamics. 

\section*{Acknowledgments}
This work was supported by DFG in the framework of SFB 910 and by the Russian Science Foundation (grant \#~16-12-10175).

\clearpage

\section*{References}
% \bibliography{iopart-num}

\end{document}